\documentclass[11pt,a4paper]{article}
\pdfoutput=1
\usepackage{jcappub}

\hypersetup{colorlinks,bookmarksopen,bookmarksnumbered,citecolor=blue,
linkcolor=black,pdfstartview=FitH,urlcolor=blue}

\usepackage{graphicx}
\usepackage{amsmath}
\usepackage{amssymb}

\newcommand{\be}{\begin{equation}}
\newcommand{\ee}{\end{equation}}
\newcommand{\beq}{\begin{equation}}
\newcommand{\eeq}{\end{equation}}

\newcommand{\vect}[1]{\boldsymbol{\rm #1}}

\newcommand{\vv}{\textbf{v}}

\makeatletter
\renewcommand{\fnum@table}{\textbf{\tablename~\thetable}}
\renewcommand{\fnum@figure}{\textbf{\figurename~\thefigure}}
\makeatother

\title{
Halo-independent methods for inelastic dark matter scattering}

\def\ific{Departamento de Fisica Teorica, and IFIC, Universidad de Valencia-CSIC\\
Edificio de Institutos de Paterna, Apt. 22085, 46071 Valencia, Spain}
\def\mpi{Max-Planck-Institut f{\"u}r Kernphysik,\\ 
Saupfercheckweg 1, 69117 Heidelberg, Germany}
\def\Cincy{Department of Physics, University of Cincinnati,\\ 
Cincinnati, Ohio 45221,USA}

\author[a]{Nassim Bozorgnia,}
\author[b]{Juan Herrero-Garcia,}
\author[a]{Thomas Schwetz}
\author[c]{and Jure Zupan}

\affiliation[a]{\mpi}
\affiliation[b]{\ific}
\affiliation[c]{\Cincy}

\emailAdd{bozorgnia@mpi-hd.mpg.de}
\emailAdd{juan.a.herrero@uv.es}
\emailAdd{schwetz@mpi-hd.mpg.de}
\emailAdd{jure.zupan@cern.ch}

\abstract{

We present halo-independent methods to analyze the results of dark matter
direct detection experiments assuming inelastic scattering. We focus
on the annual modulation signal reported by DAMA/LIBRA and present
three different halo-independent tests.  First, we compare it to the
upper limit on the unmodulated rate from XENON100 using (a) the
trivial requirement that the amplitude of the annual modulation has to
be smaller than the bound on the unmodulated rate, and (b) a bound on
the annual modulation amplitude based on an expansion in the Earth's
velocity. The third test uses the special predictions of the signal
shape for inelastic scattering and allows for an internal consistency
check of the data without referring to any astrophysics. We conclude
that a strong conflict between DAMA/LIBRA and XENON100 in the
framework of spin-independent inelastic scattering can be established
independently of the local properties of the dark matter halo.}

\arxivnumber{1305.3575}

\begin{document}
\maketitle

\section{Introduction}
\label{sec:introduction}

If dark matter (DM) is a ``Weakly Interacting Massive Particle''
(WIMP) it may induce an observable signal in underground detectors by
depositing a tiny amount of energy after scattering with a nucleus in the detector material
\cite{Goodman:1984dc}. Many experiments are currently exploring this
possibility and delivering a wealth of data. Among them is the
DAMA/LIBRA experiment~\cite{Bernabei:2010mq} (DAMA for short) which
reports the striking signature of an annual modulation of the signal
in their NaI scintillator detector, with a period of one year and a
maximum around June 2nd with very high statistical significance. Such
an effect is expected for DM induced events because the velocity of
the detector relative to the DM halo changes due to the Earth's
rotation around the Sun \cite{Drukier:1986tm, Freese:1987wu}.

Assuming elastic spin-independent interactions the DAMA modulation
signal is in strong tension with constraints on the total DM
interaction rate from other experiments \cite{Ahmed:2009zw,
  Ahmed:2010wy, Angle:2011th, Aprile:2012nq}.
  This problem can be
alleviated by considering inelastic scattering
\cite{TuckerSmith:2001hy}, where the DM particle $\chi$ up-scatters to
an excited state $\chi^*$ with a mass difference $\delta = m_{\chi^*}
- m_\chi$ comparable to the kinetic energy of the incoming particle,
which is typically $\mathcal{O}(100)$  keV. Under this hypothesis scattering off the
heavy iodine nucleus is favoured compared to the relatively light
sodium in the NaI crystal used in DAMA. Furthermore, the relative
strength of the modulation signal compared to the unmodulated rate can
be enhanced.\footnote{The possibility to use inelastic scattering to
  reconcile the event excess observed in
  CRESST-II~\cite{Angloher:2011uu} with other bounds has been
  discussed in \cite{Kopp:2011yr}. Here we do not follow this
  hypothesis and focus on the DAMA modulation signal.} Nevertheless,
under specific assumptions for the DM halo---typically a Maxwellian
velocity distribution---also the inelastic scattering explanation of
the DAMA signal is in tension with the bounds from
XENON100~\cite{Aprile:2011ts} and CRESST-II~\cite{Angloher:2011uu},
see e.g., \cite{Chang:2008gd, SchmidtHoberg:2009gn, Kopp:2009qt,
  Arina:2012dr}. Below we show that this conclusion can be confirmed
in a halo-independent way. Let us mention that any explanation of the
DAMA signal based on iodine scattering is disfavoured also by KIMS
results~\cite{Kim:2012rza}, since their 90\%~CL upper bound on the DM
scattering rate on CsI is already somewhat lower than the size of the
modulation amplitude observed in DAMA. This tension is completely
independent of astrophysics as well as particle physics as long as
scattering happens on iodine. 

For typical inelastic scattering explanations of DAMA the mass
splitting between the two DM states is chosen such that the minimal
velocity, $v_m$, required to deposit the threshold energy in the
detector is already close to the galactic escape velocity, $v_{\rm
  esc}$. Only DM particles with velocities in the interval, $v \in [v_{\rm esc} -
\Delta v, v_{\rm esc}]$, contribute, where $\Delta v$ is the range of minimal velocities  probed by the experiment and is
comparable to the Earth's velocity around the Sun, $v_e \approx
30$~km/s. In this case the DM direct detection experiment probes the
tails of the DM velocity distribution, where halo-substructures such as streams or debris flows are expected. The results are thus quite sensitive
to the exact history of the Milky Way halo, mergers, etc,
and significantly depend on the halo properties, see
e.g.~\cite{MarchRussell:2008dy, Lisanti:2010qx}. Therefore it is important to develop
astrophysics-independent methods to evaluate whether the above
conclusion on the disagreement of the DAMA signal with other bounds is
robust with respect to variations of DM halo properties. 

An interesting method to compare signals and/or bounds from different
experiments in an astrophysics independent way has been proposed in
Refs.~\cite{Fox:2010bz, Fox:2010bu}. This so-called $v_m$-method has been applied in
various recent studies for elastic scattering, see e.g.,
\cite{McCabe:2011sr, Frandsen:2011gi, Gondolo:2012rs,
  HerreroGarcia:2012fu, Frandsen:2013cna, DelNobile:2013cta}. The
generalization of this method to inelastic scattering involves some
complications which we are going to address in detail below. 

In part of our analyses we will also use the fact that $v_e$ is small compared to all other typical velocities
in the problem. One can then derive astrophysics-independent bounds
on the annual modulation signal \cite{HerreroGarcia:2011aa} by
expanding in $v_{\rm e}$ and relating the ${\mathcal O}( v_e^0)$ and ${\mathcal O}(v_e)$ terms
in a halo-independent way. In
\cite{HerreroGarcia:2012fu,HerreroGarcia:2011aa} the expansion was
applied to the case of elastic DM scattering with DM masses of order
10~GeV, where the expansion is expected to be well-behaved, and it has
been shown that for elastic scattering a strong tension between DAMA
and constraints from other experiments can be established independent
of the details of the DM halo. In the following we will generalize this
type of analysis to the case of inelastic scattering, where special
care has to be taken about whether the expansion in $v_e$ remains
well-behaved.

Below we will present three different tests for the
consistency of the inelastic scattering interpretation
of the DAMA signal, focusing on the tension with the bound from 
XENON100~\cite{Aprile:2012nq}:
\begin{itemize}
\item
the ``trivial bound'' obtained by the requirement that the amplitude
of the annual modulation has to be smaller than the unmodulated rate,
\item
the bound on the annual modulation signal based on the expansion of
the halo integral in $v_e$, and
\item
a test based on the predicted shape of the signal in the case of
inelastic scattering which we call the ``shape test'' in the following.
\end{itemize}

The paper is structured as follows.  We fix basic notation in
Sec.~\ref{sec:notation}. In Sec.~\ref{sec:bound} we discuss the bound
on the annual modulation amplitude derived in
\cite{HerreroGarcia:2011aa}. By identifying the relevant expansion
parameter we point out its limitations in the case of inelastic
scattering.  In Sec.~\ref{sec:inelastic} we develop halo-independent
methods for inelastic scattering, focusing on the tension between DAMA
and XENON100, and apply the three different types of tests mentioned
above. Conclusions are presented in Sec.~\ref{sec:conclusions}.

\section{Notation}
\label{sec:notation}

\begin{figure}
\begin{center}
  \includegraphics{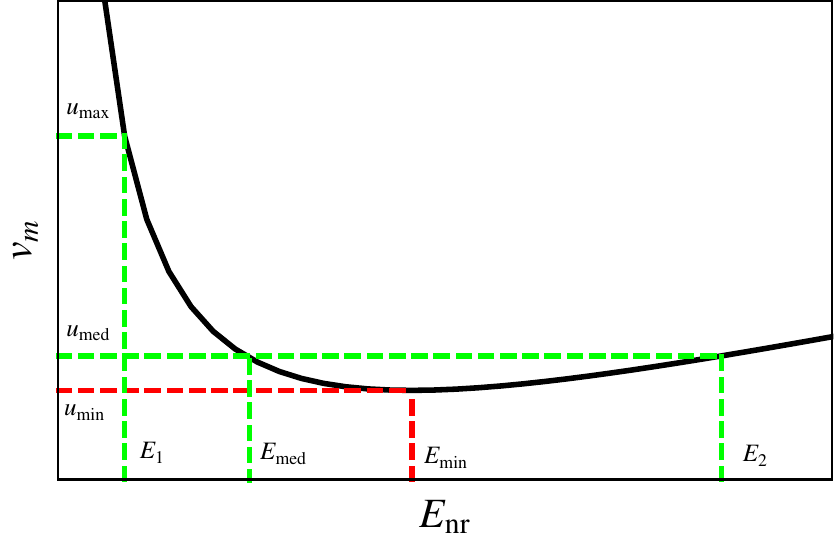}
  \caption{\label{fig:vm-inelastic} $v_m$ as a function of $E_{nr}$ for the case of inelastic scattering for some arbitrary $\delta>0$. 
  }
\end{center}
\end{figure}

The differential rate in events/keV/kg/day for DM $\chi$ to scatter
off a nucleus $(A,Z)$ and deposit the nuclear recoil energy
$E_{nr}$ in the detector is
\beq \label{rate}
{R}(E_{nr},t) = \frac{\rho_\chi}{m_\chi} \frac{1}{m_A}\int_{v>v_{m}}d^3 v \frac{d\sigma_A}{d{E_{nr}}} v f_{\rm det}(\vect v, t).
\eeq
Here $\rho_\chi \simeq 0.3 \, {\rm GeV/cm}^3$ is the local DM density, $m_A$
and $m_\chi$ are the nucleus and DM masses, $\sigma_A$ the DM--nucleus
scattering cross section and $\vect v$ the 3-vector relative velocity
between DM and the nucleus, while $v\equiv |\vect{v}|$. For a DM particle to
deposit a recoil energy $E_{nr}$ in the detector, a minimal velocity $v_{m}$ is
required, restricting the integral over velocities in Eq.~\eqref{rate}. 
For inelastic scattering we have
\beq
v_m=\sqrt{\frac{1}{2 m_A E_{nr}}}\left(\frac{m_A E_{nr}}{\mu_{\chi A}} + \delta \right),
\label{v-inelastic}
\eeq
where $\mu_{\chi A}$ is the reduced mass of the DM-nucleus system, and $\delta$
is the mass splitting between the two dark matter states. Note that
for each value of $E_{nr}$ there is a corresponding $v_m$ while the converse
is not always true. Certain values of $v_m$ correspond to two values of
$E_{nr}$, others maybe to none. This is illustrated in 
Fig.~\ref{fig:vm-inelastic}, where we plot $v_m$ as a function of
$E_{nr}$ for some arbitrary $\delta>0$. 

The particle physics enters in Eq.~\eqref{rate} through the differential cross section. For the standard  spin-independent and spin-dependent scattering  the differential cross section is 
\begin{align}
  \frac{d\sigma_A}{dE_{nr}} = \frac{m_A}{2\mu_{\chi A}^2 v^2} \sigma_A^0 F^2(E_{nr}) \,, 
  \label{eq:dsigmadE}
\end{align}
where $\sigma_A^0$ is the total DM--nucleus scattering cross section
at zero momentum transfer, and $F(E_{nr})$ is a form factor. We focus
here on spin-independent inelastic scattering. We also assume that DM couples with the same strength to protons and neutrons ($f_p=f_n$). Relaxing this assumption
 does not change the conclusions since DM particles scattering on Xe and I have very similar dependence on $ f_n/f_p$ (cf. Fig. 5 of \cite{Kopp:2011yr}).  The
astrophysics dependence enters in Eq.~\eqref{rate} through the DM velocity
distribution $f_{\rm det}(\vect{v}, t)$ in the detector rest
frame. Defining the halo integral
\beq\label{eq:eta} 
\eta(v_m, t) \equiv \int_{v > v_m} d^3 v \frac{f_{\rm det}(\vect{v}, t)}{v} \,,
\eeq
the event rate is given by
\beq\label{eq:Rgamma}
R(E_{nr}, t) = C \, F^2(E_{nr}) \, \eta(v_m, t) 
\qquad\text{with}\qquad
C = \frac{\rho_\chi \sigma_A^0}{2 m_\chi \mu_{\chi A}^2}.
\eeq
The coefficient $C$ contains the particle physics dependence, while
$\eta(v_m,t)$ parametrizes the astrophysics dependence. The halo
integral $\eta(v_m,t)$ is the basis for the astrophysics independent
comparison of experiments~\cite{Fox:2010bz, Fox:2010bu} and we will
make extensive use of it below.

\section{Bound on the annual modulation amplitude from the expansion in $v_e$}
\label{sec:bound}

In Ref.~\cite{HerreroGarcia:2011aa} some of us have derived an upper
bound on the annual modulation amplitude in terms of the unmodulated
rate. Here we briefly review the idea and generalize the bound to the
case of inelastic scattering, where special care has to be taken about
the validity of the expansion.

The DM velocity distribution in the rest frame of the Sun,
$f(\vect{v})$, is related to the distribution in the detector rest
frame by $f_{\rm det}(\vv, t) = f(\vv + \vv_e(t))$.  The basic
assumption of \cite{HerreroGarcia:2011aa} is that $f(\vect{v})$ is
constant in time on the scale of 1 year and is constant in space on
the scale of the size of the Sun--Earth distance. These are very weak
requirements, called ``Assumption 1'' in \cite{HerreroGarcia:2011aa},
which are expected to hold for a wide range of possible DM
  halos. Those assumptions would be violated if a few DM substructures
  of $\sim$1~AU in size would dominate the local DM distribution. The
  smallest DM substructures in typical WIMP scenarios can have masses
  many orders of magnitude smaller than $M_\odot$,
  e.g.~\cite{Bringmann:2009vf}. Based on numerical simulations it is
  estimated in \cite{Diemand:2005vz} that Earth mass DM substructures
  with sizes comparable to the solar system are stable against
  gravitational disruption, and on average one of them will pass through the solar system
  every few thousand years, where such an encounter would last about
  50 years. Those considerations suggest that Assumption~1 is well
  satisfied. Let us stress that typical DM streams or debris flows
  \cite{Kuhlen:2012fz} which may dominate the DM halo at high
  velocities (especially relevant for inelastic scattering) are expected to be many orders of magnitude larger than
  1~AU, and the relevant time scales are much larger than 1~yr, and hence they
  fulfill our assumptions, see e.g.~\cite{Freese:2012xd} and references therein.

Under this assumption the only time
dependence is due to the Earth's velocity $\vect{v}_e(t)$, which we write as
\cite{Gelmini:2000dm}
\beq\label{eq:ve}
\vect{v}_e(t) = v_e [\vect{e}_1 \sin\lambda(t) - \vect{e}_2\cos\lambda(t) ],
\eeq
with $v_e=29.8$ km/s, and $\lambda(t)=2\pi(t-0.218)$ with $t$ in units
of 1 year and $t=0$ at January 1st, while $\vect e_1 =
(-0.0670,0.4927,-0.8676)$ and $\vect e_2 =(-0.9931,-0.1170,0.01032)$
are orthogonal unit vectors spanning the plane of the Earth's orbit
which at this order can be assumed to be circular. The DM velocity
distribution in the galactic frame is connected to the one in the rest
frame of the Sun by $f(\vect{v}) = f_{\rm gal}(\vect{v} + \vect{v}_{\rm sun})$, 
with $\vect v_{\rm sun} \approx (0,220,0) \, {\rm km/s} + \vect v_{\rm
  pec}$ and $\vect v_{\rm pec} \approx (10,13,7)$~km/s the peculiar
velocity of the Sun. We are using galactic coordinates where $x$
points towards the galactic center, $y$ in the direction of the
galactic rotation, and $z$ towards the galactic north, perpendicular
to the disc. As shown in \cite{Green:2003yh}, Eq.~\eqref{eq:ve}
provides an excellent approximation to describe the annual modulation
signal.  

Using the fact that $v_e$ is small compared to other relevant
velocities, one can expand the halo integral Eq.~\eqref{eq:eta} in
powers of $v_e$. At zeroth order one obtains
\begin{equation}\label{eq:eta0}
  \eta_0(v_m) = \int_{v > v_m} d^3 v \frac{f(\vect{v})}{v} \,,  
\end{equation}
which is responsible for the unmodulated (time averaged) rate up to
terms of order $v_e^2$. The first order terms in $v_e$ lead to the
annual modulation signal, which due to Eq.~\eqref{eq:ve} will have a
pure sinusoidal shape, such that
\begin{equation}
  \eta(v_m, t) = \eta_0(v_m) + A_\eta(v_m) \cos 2\pi[t - t_0(v_m)] + \mathcal{O}(v_e^2) \,,
\end{equation}
where the amplitude of the annual modulation, $A_\eta(v_m)$, is of first order in $v_e$.

In \cite{HerreroGarcia:2011aa} it has been shown that under the above stated ``Assumption 1''
the modulation amplitude is bounded as
\begin{align}\label{ineq-previous}
  A_\eta(v_m) < v_e 
  \left[- \frac{d\eta_0}{dv_m} + \frac{\eta_0}{v_m} - \int_{v_m} dv \frac{\eta_0}{v^2} \right]\,.
\end{align}
From Eq.~\eqref{eq:eta0} it is clear that $\eta_0$ is a positive
decreasing function, i.e., $d\eta_0/dv_m < 0$. As mentioned above, in
the case of inelastic scattering typically only a small range in
minimal velocities $v_m$ is probed. We denote this interval by
$[u_{\rm min},u_{\rm max}]$ with $\Delta v = u_{\rm max} - u_{\rm min}$. The boundaries of this
interval are determined by the threshold of the detector on one
side and by the galactic escape velocity or the nuclear form factor
suppression on the other side. For inelastic scattering $\Delta
v$ is small. It will thus be convenient to integrate the inequality \eqref{ineq-previous} over
the interval $[u_{\rm min},u_{\rm max}]$.  By changing the order of integrations of
the double integral we find
\begin{align}
 \int_{u_{\rm min}}^{u_{\rm max}} dv A_\eta(v) & < v_e \left[\eta_0(u_{\rm min}) - \eta_0(u_{\rm max}) 
   +  u_{\rm min} \int_{u_{\rm min}}^{u_{\rm max}} dv \frac{\eta_0}{v^2}
   -  \Delta v \int_{u_{\rm max}} dv \frac{\eta_0}{v^2} \right]\nonumber\\
  & <  v_e \left[ \eta_0(u_{\rm min}) 
   +  u_{\rm min} \int_{u_{\rm min}}^{u_{\rm max}} dv \frac{\eta_0}{v^2}  \right].
\end{align}
Integrating again over $u_{\rm min}$ we obtain
\begin{align}
 \int_{u_{\rm min}}^{u_{\rm max}} dv A_\eta(v)(v-u_{\rm min}) & < 
   \frac{v_e}{2}\int_{u_{\rm min}}^{u_{\rm max}} dv \eta_0
   \left(3 - \frac{u_{\rm min}^2}{v^2}\right) \label{eq:bound-mod-alt}\\
 & < 
 \frac{v_e}{2} \left(3 - \frac{u_{\rm min}^2}{u_{\rm max}^2}\right)    
   \int_{u_{\rm min}}^{u_{\rm max}} dv \eta_0 \,.\label{eq:bound-mod}
\end{align}
Hence we obtained a bound on the integral of the annual modulation in
terms of an integral of the unmodulated rate at first order in
$v_e$.\footnote{Below we will use the bound \eqref{eq:bound-mod} for the
  numerical analysis since this will allow for easy comparison with
  the ``trivial bound'' discussed later. The numerical difference between the bounds
  using \eqref{eq:bound-mod} or \eqref{eq:bound-mod-alt} is small, typically less than
  10\%.}  This bound receives no corrections at order ${\mathcal O}(v_e^2)$ and
hence is valid up to 
(but not including) terms of order
${\mathcal O}(v_e^3)$ \cite{BHSZ}. In applying Eq.~\eqref{eq:bound-mod} we use only
an upper bound, $\eta_{\rm bnd}$ on the unmodulated signal $\eta_0$, allowing also for the presence of
background. However, we assume that $A_\eta$ is background free, i.e.,
the background shows no annual modulation.

Let us define the average over the velocity interval by 
\begin{equation}
  \langle X \rangle = \frac{1}{\Delta v} \int_{u_{\rm min}}^{u_{\rm max}} dv X(v) \,.
\end{equation}
Estimating $\int_{u_{\rm min}}^{u_{\rm max}} dv A_\eta(v)(v-u_{\rm min})
\sim \Delta v \int_{u_{\rm min}}^{u_{\rm max}} dv A_\eta(v)$ and neglecting
$\mathcal{O}(1)$ coefficients we obtain from Eq.~\eqref{eq:bound-mod}
\begin{equation}
  \langle A_\eta \rangle \lesssim \frac{v_e}{\Delta v}
  \langle \eta_0 \rangle \,.
\end{equation}
This shows that the expansion parameter in deriving the bound \eqref{eq:bound-mod} is $v_e/\Delta v$. In
contrast to expressions like $v_e/u_{\rm min}$ which are always small, the
ratio $v_e/\Delta v$ can become of order one, in particular for
inelastic scattering.

\begin{figure}
\begin{center}
  \includegraphics[height=240pt]{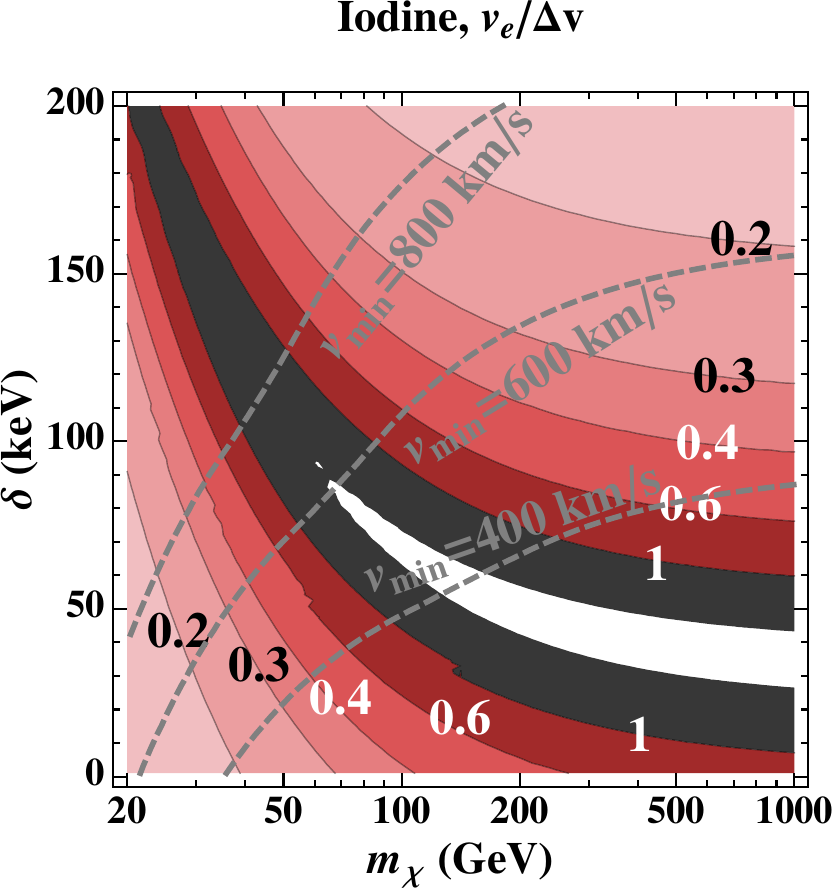}
  \caption{\label{fig:vevDelta} Contours of the ratio $v_e/\Delta v$
    as a function of DM mass $m_\chi$ and mass splitting $\delta$ for
    scattering on iodine in DAMA, assuming that $\Delta v$ is the
    overlap between the velocity ranges corresponding to the DAMA
    $[2,4]$~keVee energy range for scattering on iodine and the
    XENON100 $[6.61,43.04]$ keV range. Dashed curves show contours of
    constant $v_m = 400, 600, 800$~km/s, with $v_m$ being the minimal
    velocity corresponding to the above energy interval for the given
    $\delta$ and $m_\chi$. 
    } 
\end{center}
\end{figure}

As an example we show in Fig.~\ref{fig:vevDelta} the ratio $v_e/\Delta
v$ as a function of the DM mass $m_\chi$ and the inelasticity
parameter $\delta$. Elastic scattering is recovered for $\delta =
0$. The velocity interval $\Delta v$ is chosen having in mind a
possible explanation for DAMA. The signal in DAMA is predominantly in
the energy region $[2,4]$~keVee. As explained in
Section~\ref{sec:inelastic} below, we take for $\Delta v$ the overlap
between the velocity ranges corresponding to the DAMA $[2,4]$~keVee
interval for scattering on iodine and the XENON100 $[6.61,43.04]$~keV
interval. From Fig.~\ref{fig:vevDelta} one sees that the expansion
parameter $v_e/\Delta v\gtrsim 1$ for a non-negligible part of the
parameter space of DM masses $m_\chi$ and mass splittings
$\delta$. For those values of $m_\chi$ and $\delta$ the bound
Eq.~\eqref{eq:bound-mod} does not apply. Still, in a significant part
of the parameter space $v_e/\Delta v$ is sufficiently small such that
the expansion can be performed. In particular, we observe from the
figure that for elastic scattering ($\delta = 0$) and $m_\chi \lesssim
50$ the expansion parameter is small, justifying the approach of
Ref.~\cite{HerreroGarcia:2012fu}. 

\section{Halo-independent tests for inelastic scattering}
\label{sec:inelastic}

In this section we present three different halo-independent tests of the tension
between the DAMA annual modulation signal \cite{Bernabei:2010mq} and
the bound from XENON100~\cite{Aprile:2012nq} in the framework of
inelastic scattering. The tests are presented in the following order. First, we present the shape test which is  a test based on the predicted shape of the signal.  Second, we present the bound on the annual modulation signal from Eq.~\eqref{eq:bound-mod}. Third, we present the trivial bound which is based on the fact that the amplitude of the annual modulation must be smaller than the unmodulated rate.

The $v_m$ method \cite{Fox:2010bz, Fox:2010bu}
to compare different experiments like DAMA and XENON100 requires to
translate the physical observations in nuclear recoil energy $E_{nr}$
into $v_m$ space using Eq.~\eqref{v-inelastic}. Then experiments can
be directly compared based on the halo integral $\eta(v_m)$ or
inequalities such as Eq.~\eqref{eq:bound-mod}
\cite{HerreroGarcia:2012fu}. However, for inelastic scattering this
involves some complications. The reason is that in inelastic
scattering each minimal velocity $v_m$ can correspond to up to two
values of $E_{nr}$, depending on the values of $m_\chi$ and
$\delta$. This has to be taken into account when translating an
observation at a given $E_{nr}$ into $v_m$, since the relation between
them is no longer unique (as it is for elastic scattering).  Solving
Eq.~\eqref{v-inelastic} for $E_{nr}$, one obtains two solutions
$E_{\pm}$ as a function of $v_m$,
\beq
E_{\pm}=\left(\frac{\mu_{\chi A}}{m_A}\right)\left[ (\mu_{\chi A} v_m^2 -\delta) \pm v_m
  \sqrt{\mu_{\chi A} (\mu_{\chi A} v_m^2 - 2 \delta)} \right]. 
\label{E_pm}
\eeq
There is a minimal value of $v_m$ given by $\sqrt{2\delta/\mu_{\chi A}}$ at an
energy $E_{\rm min} = \mu_{\chi A}\delta/M_A$.

Let us consider the following situation, having in mind DAMA: we have
a region $[E_1, E_2]$ in nuclear recoil energy where the modulation
amplitude is non-zero. We assume that $E_1$ is the threshold energy of
the detector. When mapped into $v_m$ space according to
Eq.~\eqref{v-inelastic} we obtain that the whole interval $[E_1, E_2]$
is mapped into a small region in $v_m$, between $u_{\rm min}$ and
$u_{\rm max}$ with $u_{\rm max} -u_{\rm min} \ll u_{\rm min}$, where
$u_{\rm min}$ and $u_{\rm max}$ are the minimum and maximum values of
$v_m$ in the $[E_1, E_2]$ interval, respectively. For the special case
plotted in Fig.~\ref{fig:vm-inelastic}, $u_{\rm min}=v_m(E_{\rm
  min})$, and $u_{\rm max}=v_m(E_1)$.  In general, depending on the
shape of $v_m$ as a function of $E_{nr}$ in the interval $[E_1 ,
  E_2]$, $u_{\rm max}$ may either be $v_m(E_1)$ (as in the case shown
in Fig.~\ref{fig:vm-inelastic}) or $v_m(E_2)$. Furthermore, in cases
where $E_{\rm min}$ falls outside of the interval $[E_1, E_2]$,
$u_{\rm min}$ will either be $v_m(E_1)$ or $v_m(E_2)$. We will not
discuss all these cases here explicitly, but as an example focus on
the case shown in Fig.~\ref{fig:vm-inelastic}.

To compute the bound in Eq.~\eqref{eq:bound-mod} using DAMA
data, we need to numerically compute integrals such as
$\int_{u_{\rm min}}^{u_{\rm max}} d v h(v) A_\eta^{\rm obs}(v)$ where $h(v) = v - u_{\rm min}$
is specified in Eq.~\eqref{eq:bound-mod} (we leave it general
here to apply the same formalism also to the bound in
Eq.~\eqref{eq:trivial} discussed later on, where $h(v) = 1$) and
$A_\eta^{\rm obs}(v)$ is the observed amplitude of the annual
modulation in units of events/kg/day/keV. In order to compute
those integrals we have to consider the functional relation between
$v_m$ and $E_{nr}$ in the relevant interval $[E_1,E_2]$. Let us
discuss for instance the situation depicted in
Fig.~\ref{fig:vm-inelastic}. In this case we have
\begin{align}
\int_{u_{\rm min}}^{u_{\rm max}} d v h(v) \tilde{A}_\eta^{\rm obs}(v)&=\int_{u_{\rm min}}^{u_{\rm med}} d v h(v) \tilde{A}_\eta^{\rm obs}(v)+\int_{u_{\rm med}}^{u_{\rm max}} d v h(v) \tilde{A}_\eta^{\rm obs}(v)\nonumber\\
&=\int_{u_{\rm min}}^{u_{\rm med}} d v h(v) \tilde{A}_\eta^{\rm obs}(v) + \int_{E_{\rm med}}^{E_1} dE_{nr} \frac{d v}{d E_{nr}} h(E_{nr}) \tilde{A}_\eta^{\rm obs}(v).
\label{int-A-Bin1}
\end{align}
Here, $u_{\rm med}=v_m(E_2)$ and $E_{\rm med}=E_{-}(u_{\rm med})$. 
The integrals can be written as a sum of several integrals which are
evaluated over energy bins, as given by the DAMA binning. We take four bins of equal size in the [2, 4] keVee range for the DAMA data. In each bin
we write \cite{HerreroGarcia:2012fu}
\beq\label{eq:A-tilde}
\tilde{A}_\eta^{\rm obs} (v_i)=\frac{A_{i}^{\rm obs} q_I}{A_{\rm I}^2 F_{\rm I}^2 (E_{nr}) f_{\rm I}} \,,
\eeq
where the index $i$ labels energy bins, $q_I$ is the iodine quenching
factor for which we take $q_I=0.09$\footnote{For DM masses that we consider one can safely neglect scattering on sodium. Note also that the channeling fraction of iodine in NaI is likely to be very
small and can be neglected~\cite{Bozorgnia:2010xy}.}, $F_I (E_{nr})$ is the Helm form
factor for iodine, and $f_I=m_{\rm I}/(m_{\rm Na}+m_{\rm I})$. In each
energy bin we assume $A_{i}^{\rm obs}$ is constant, and thus in each
bin we numerically integrate $(d v/dE_{nr})h(E_{nr})/F_{\rm
  I}^2(E_{nr})$ over the bin width.

\begin{figure}
\begin{center}
  \includegraphics[height=230pt]{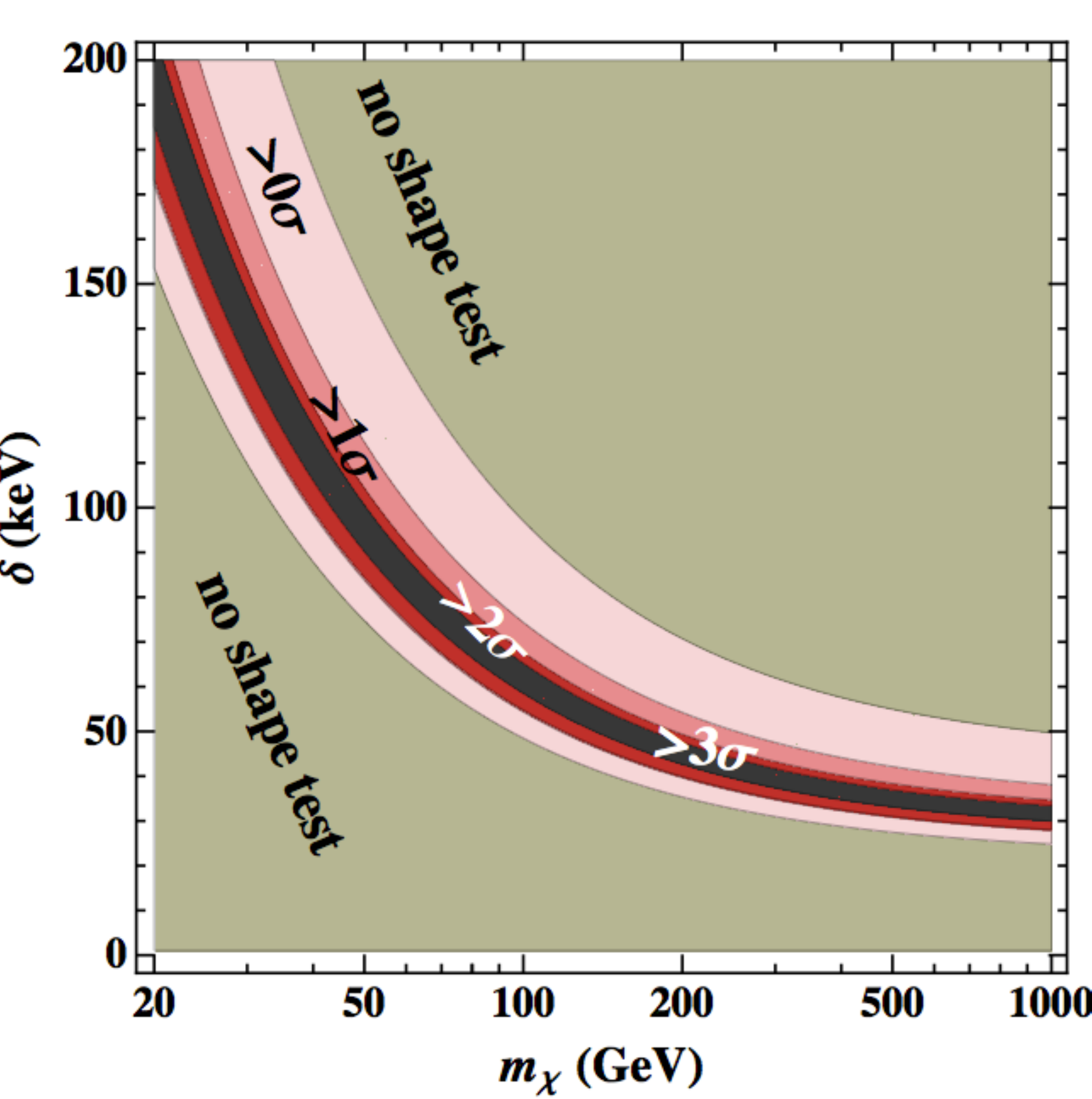}
  \caption{\label{fig:inelast-integrals} The DM exclusion regions (red bands, with significance as denoted) 
   that follow from the internal consistency shape test 
 for DAMA data, see Eq.~\eqref{eq:difference}. In the gray region denoted by ``no shape test''
    there is a one-to-one correspondence between $E_{nr}$ and $v_m$
    since $E_{\rm min}$ lies outside the relevant energy interval
    $[E_1,E_2]$ and therefore the shape test cannot be applied. 
    }
\end{center}
\end{figure}

For the first integral on the r.h.s.\ of Eq.~\eqref{int-A-Bin1} there
is an ambiguity, since the interval $[u_{\rm min}, u_{\rm med}]$
corresponds to two regions in energy: $[E_{\rm med}, E_{\rm min}]$ or
$[E_{\rm min}, E_2]$. If the inelastic DM hypothesis under
consideration is correct, both energy intervals should give the same
value of the integral. We can use this observation to test the hypothesis that the signal is due to inelastic DM scattering
by requiring that the two integrals agree within
experimental errors.  In the
following, we will call this the ``shape test''.  Let us denote the integrals corresponding to the two
energy intervals by $I_a$ and $I_b$ and their experimental errors by
$\sigma_a$ and $\sigma_b$, correspondingly. In
Fig.~\ref{fig:inelast-integrals} we show the difference weighted by
the error as obtained from DAMA data:
\beq\label{eq:difference}
\frac{|I_a - I_b|}{\sqrt{\sigma_a^2 + \sigma_b^2}} \,.
\eeq
We observe that a strip in the parameter space in $\delta$ and
$m_\chi$ is already excluded by this requirement at more than
$3\sigma$ in a completely halo-independent way, just requiring a
spectral shape of the signal consistent with the inelastic scattering
hypothesis. In cases where the two values are consistent within errors
we use for the integral the weighted average of the two values. In 
Fig.~\ref{fig:inelast-integrals}, $I_a$ and $I_b$ are evaluated for the choice of
 $h(v)=v-u_{\rm min}$. The shape test is only slightly different for $h(v)=1$ which is the case 
 for the trivial bound explained later in Eq.~\eqref{eq:trivial}.

\bigskip

In order to evaluate the r.h.s.\ of the inequality in
Eq.~\eqref{eq:bound-mod},
we need to calculate an integral over the experimental upper bound
$\tilde{\eta}_{\rm bnd}(v_m)$ on the unmodulated signal, with
\begin{equation}\label{eq:eta-tilde}
  \tilde\eta(v_m) \equiv \frac{\sigma_p \rho_\chi}{2 m_\chi \mu^2_{\chi p}} \eta_0(v_m) \,,
\end{equation}
where $\sigma_p$ is the cross section on a nucleon and $\mu_{\chi p}$
is the DM--nucleon reduced mass. $\tilde\eta$ has units of
events/kg/day/keV.  In using Eqs.~\eqref{eq:A-tilde} and
\eqref{eq:eta-tilde} we have assumed an $A^2$ dependence of the
scattering cross section on the nucleus with mass number $A$. We use
the method discussed in Ref.~\cite{HerreroGarcia:2012fu} (see also
\cite{Fox:2010bz}) to evaluate $\tilde{\eta}_{\rm bnd}(v_m)$ for the
inelastic case. Namely, we use the fact that $\tilde{\eta}(v_m)$ is a
falling function, and that the minimal number of events is obtained
for $\tilde{\eta}$ constant and equal to $\tilde{\eta}(v_m)$ up to
$v_m$ and zero for larger values of $v_m$. Therefore, for a given
$v_m$ we have a lower bound on the predicted number of events in an
interval of observed energies $[E_1, E_2]$, $N^{\rm
  pred}_{[E_1,E_2]}>\mu(v_m)$ with
\beq
\mu(v_m)=MT A^2 \tilde{\eta}(v_m) \int_{E_-}^{E_+} dE_{nr} F_A^2(E_{nr}) G_{[E_1,E_2]}(E_{nr}),
\label{mu}
\eeq
where $G_{[E_1,E_2]}(E_{nr})$ is the detector response function which
describes the contribution of events with the nuclear-recoil energy
$E_{nr}$ to the observed energy interval $[E_1, E_2]$. $M$ and $T$ are the detector mass and exposure time, respectively.
Notice that
$\mu(v_m)$ for the elastic case is given in Eq.~(10) of
Ref.~\cite{HerreroGarcia:2012fu} and in that case the integral is
computed between 0 and $E(v_m)$ which corresponds to velocities below a
fixed $v_m$. For the inelastic case, we have two solutions $E_+$ and
$E_-$ for each $v_m$, and the region in velocity space below $v_m$ is
precisely, the region $E_-<E_{nr}<E_+$.

Assuming an experiment observes $N^{\rm obs}_{[E_1,E_2]}$ events in
the interval $[E_1, E_2]$, we can obtain an upper bound on
$\tilde{\eta}(v_m)$ for a fixed $v_m$ at a confidence level CL by
requiring that the probability of obtaining $N^{\rm obs}_{[E_1,E_2]}$
events or less for a Poisson mean of $\mu(v_m)$ is equal to
$1-$CL. The upper bound obtained in this way is $\tilde{\eta}_{\rm
  bnd}(v_m)$ and can then be used in Eq.~\eqref{eq:bound-mod}
and numerically integrated over $[u_{\rm min}, u_{\rm max}]$ to
constrain the modulation amplitude. We use the data from
XENON100~\cite{Aprile:2012nq} where the $E_{nr}$ interval
$[6.61,43.04]$~keV is binned into four bins. In each bin we calculate
the probability of obtaining $N^{\rm obs}_{[E_1,E_2]}$ events or less
for a Poisson mean of $\mu(v_m)$ as described above, and then multiply
the probability of the four bins to obtain the overall probability, giving finally 
 the actually observed event distribution. Note, that since only the high energy 
range in Xenon is relevant, our results are not sensitive to uncertainties 
in the scintillation efficiency $\mathcal{L}_{\rm eff}$ at low energies. For the comparison of
the DAMA and XENON100 data using Eq.~\eqref{eq:bound-mod} we
define the $[u_{\rm min}, u_{\rm max}]$ range as the overlap between
$v_m$ spaces corresponding to the DAMA iodine $[2,4]$ keVee range and
the XENON100 $[6.61,43.04]$ keV range.\footnote{For most of the region
  in parameter space this joint interval is actually very close to
  the one coming from DAMA iodine $[2,4]$ keVee.}

\begin{figure}
\begin{center}
 \includegraphics[width=0.49\textwidth]{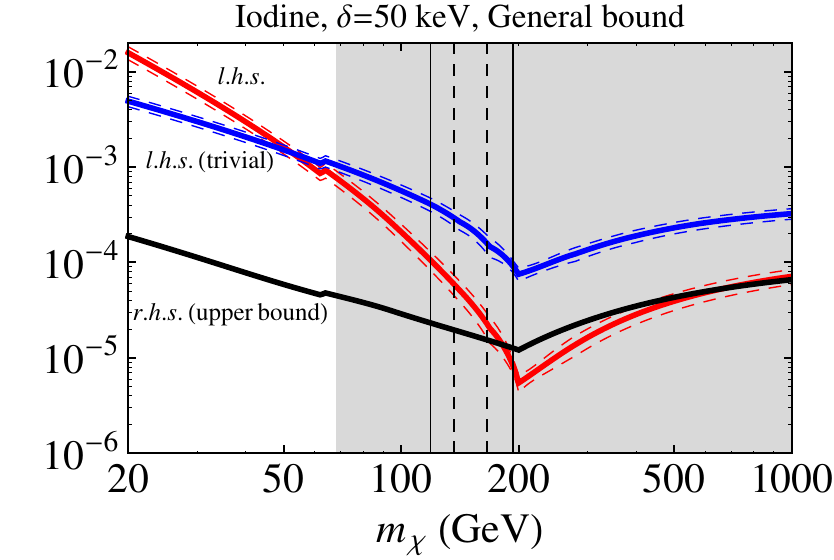}
 \includegraphics[width=0.49\textwidth]{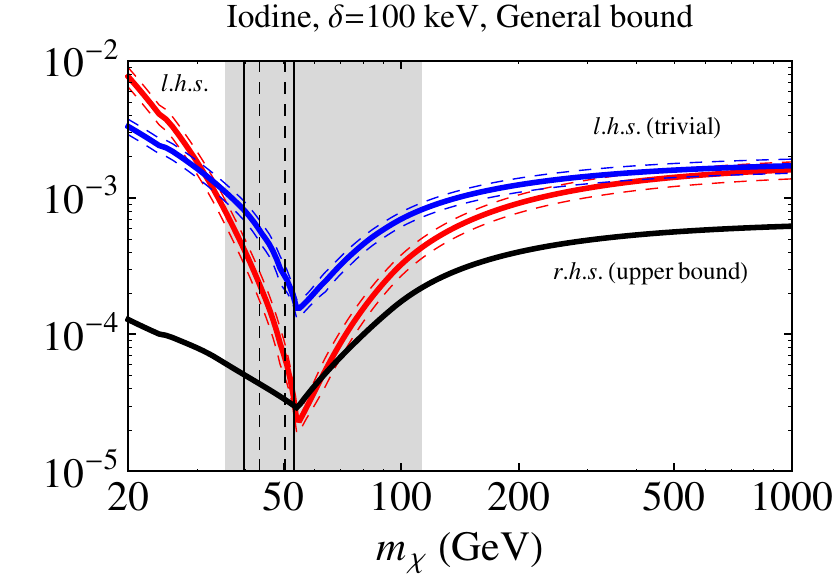}\\[5mm]
 \includegraphics[width=0.49\textwidth]{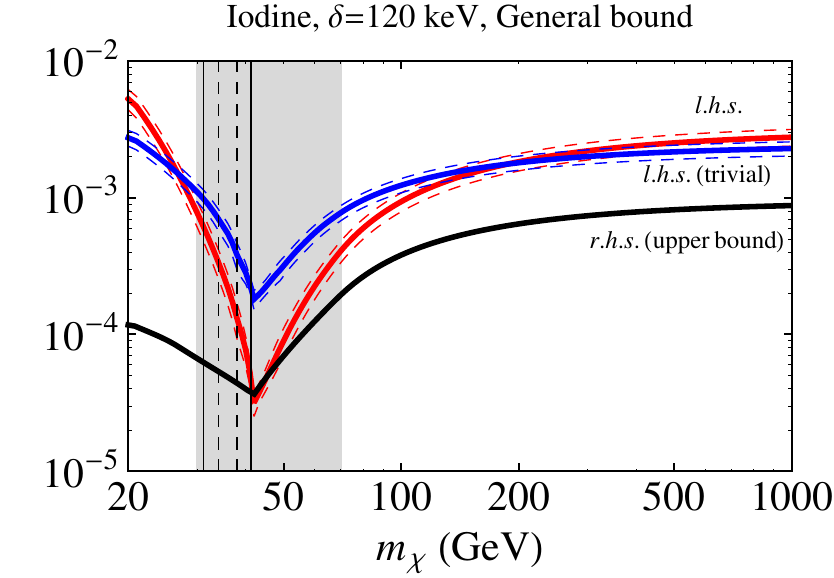}
 \includegraphics[width=0.49\textwidth]{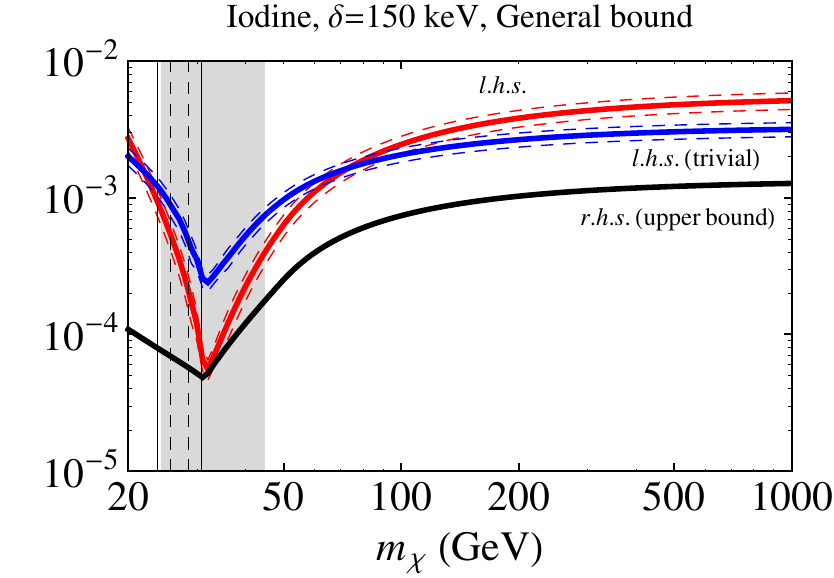}
\caption{\label{fig:Bound} The bounds from
  Eqs.~\eqref{eq:bound-mod} and \eqref{eq:trivial} for DAMA data
  as a function of $m_\chi$ for $\delta=50,100,120,150$~keV. The red
  curve labeled ``l.h.s.'' shows the integral on the l.h.s.\ of
  Eq.~\eqref{eq:bound-mod}, whereas the blue curve labeled
  ``l.h.s. (trivial)'' corresponds to the l.h.s.\ of the trivial bound
  in Eq.~\eqref{eq:trivial}. The dashed curves indicate the $1\sigma$
  error.  The black curve labeled ``r.h.s.\ (upper bound)'' is the
  same for \eqref{eq:bound-mod} and \eqref{eq:trivial} and has
  been obtained from the 3$\sigma$ limit on $\eta_{\rm bnd}$ from
  XENON100 data. The units on the vertical axis are counts/kg/day/keV~(km/s)$^2$. 
  In the gray shaded regions we have $v_e/\Delta v >
  0.7$; truncating the expansion may not be a good approximation and
  hence, the red curve should not be trusted in those regions, but instead the blue one can be used there. The
  solid (dashed) vertical lines indicate the regions where the two
  integrals relevant for the ``shape test'' differ by more than
  $2\sigma$ ($3\sigma$) according to Eq.~\eqref{eq:difference}.}
\end{center}
\end{figure}

In Fig.~\ref{fig:Bound} we show the l.h.s.\ and r.h.s.\ of the bound
from Eq.~\eqref{eq:bound-mod} in red and black, respectively, as
a function of $m_\chi$ for $\delta=50$~keV, 100~keV, 120~keV, and
150~keV. We calculate the integral over the annual modulation
amplitude in the l.h.s.\  of Eq.~\eqref{eq:bound-mod} as described above. The red dashed curves
indicate the $1\sigma$ error on the integral. The upper limit on the
r.h.s.\  of Eq.~\eqref{eq:bound-mod} is calculated from the XENON100 $3\sigma$ upper limit.
We see that in most regions of the parameter space the bound is
strongly violated, disfavoring an inelastic scattering interpretation
of the DAMA signal halo-independently. Note that DM mass enters only via $\mu_{\chi A}$, so that $\mu_{\chi A} \simeq m_A$  for  $m_\chi\gg m_A$, and  Eq.   \eqref{E_pm} becomes independent of $m_\chi$. This is what we see in Fig. \ref{fig:Bound}, where curves become flat for large $m_\chi$ and therefore the tension between XENON100 and DAMA  cannot be diminished when going to larger DM masses.

The shaded regions in Fig.~\ref{fig:Bound} are the regions where the expansion parameter $v_e/\Delta v$  is large (we take somewhat arbitrarily $v_e/\Delta v>0.7$, cf. also
Fig.~\ref{fig:vevDelta}). Hence, 
in the shaded regions 
the astrophysics independent bound on the modulation amplitude, 
Eq.~\eqref{eq:bound-mod} (the red curves in Fig.~\ref{fig:Bound}), can receive ${\mathcal O}(1)$ corrections and should not be trusted.
Interestingly, however, part of the region where the $v_e$ expansion breaks down is
disfavored by the internal consistency ``shape test'', Eq. \eqref{eq:difference}. We indicate the range in $m_\chi$ where
the two integrals differ by more than $2\sigma$ and $3\sigma$ with
vertical lines, cf.\ also Fig.~\ref{fig:inelast-integrals}.

\bigskip

Finally we consider the ``trivial bound'', which is based
on the simple fact valid for any positive function that the amplitude
of the first harmonic has to be smaller than the constant part, i.e.,
$A_\eta \le \eta_0$. To compare directly with
Eq.~\eqref{eq:bound-mod}, we can write the trivial bound as,
\begin{align}
 \frac{v_e}{2} \left(3 - \frac{u_{\rm min}^2}{u_{\rm max}^2}\right)    
   \int_{u_{\rm min}}^{u_{\rm max}} dv A_\eta(v)
< 
 \frac{v_e}{2} \left(3 - \frac{u_{\rm min}^2}{u_{\rm max}^2}\right)    
  \int_{u_{\rm min}}^{u_{\rm max}} dv \, \eta_{\rm bnd}(v) \,.
\label{eq:trivial}
\end{align}
The l.h.s.\ of this relation is shown as blue curve in
Fig.~\ref{fig:Bound} together with its $1\sigma$ error band. We
observe again strong tension with the upper bound from XENON100
data. Clearly this bound is independent of any expansion parameter and
is valid in the full parameter space. In the regions where the
expansion is expected to break down (i.e., the grey shaded regions)
this bound can be used to exclude the inelastic explanation for
DAMA. From Fig. \ref{fig:Bound} we also observe 
that in the regions
where the expansion in $v_e$ is expected to be valid the modulation bound from
Eq.~\eqref{eq:bound-mod} becomes stronger (or at least
comparable -- for large $m_\chi$) to the trivial bound.

\section{Conclusions}
\label{sec:conclusions}

Inelastic scattering \cite{TuckerSmith:2001hy} has originally been  invoked
 to reconcile the DAMA annual modulation signal with bounds
from other experiments. Using the kinematics of inelastic scattering
the annual modulation amplitude can be enhanced compared to the time
averaged rate. This is achieved at the expense of tuning the minimal
velocity probed by the experiment so that it is close to the galactic escape
velocity, which makes the signal rather sensitive to properties of the
tails of the dark matter velocity distribution. Hence it is important
to establish halo-independent methods for this scenario.

In this work we have generalized the comparison of dark matter direct detection experiments in $v_m$
space \cite{Fox:2010bz, Fox:2010bu} to the case of inelastic
scattering. This is non-trivial due to the non-unique relation between the
recoil energy and $v_m$. Turning this complication into a virtue, we
presented a consistency check based on the particular shape of the
signal for inelastic scattering which we dubbed the ``shape test'', given in Eq. \eqref{eq:difference} and Fig. \ref{fig:inelast-integrals}. In certain
regions of the parameter space the inelastic scattering hypothesis can
be excluded simply based on the energy spectrum of the modulation
signal, without referring to halo properties.

Furthermore, we have applied a bound on the annual modulation
amplitude based on an expansion of the halo integral in the Earth's
velocity $v_e$ \cite{HerreroGarcia:2011aa}. We have identified the
relevant expansion parameter to be $v_e/\Delta v$, where $\Delta v$ is
the range in minimal velocities $v_m$ probed in the experiment. For
inelastic scattering, $\Delta v$ can become of order $v_e$ 
for part of the $(m_\chi,\delta)$ parameter space, and then the bound cannot be
applied. However, in those cases one can use the ``trivial bound'',
requiring that the amplitude of the annual modulation has to be less
than the bound on the unmodulated rate.

We were able to show that XENON100 strongly disfavors an
interpretation of the DAMA modulation signal in terms of inelastic
scattering, independent of assumptions on the properties of the local
dark matter velocity distribution. Beyond the immediate problem of
interpreting the DAMA signal, the methods developed in this manuscript
will provide a valuable consistency check for an inelastic scattering
interpretation of any future dark matter signal.

In our work we have focused on spin-independent contact
  interactions, where the differential scattering cross section takes
  the form of Eq.~\eqref{eq:dsigmadE}. Our considerations generalize
  trivially to other interaction types which lead to a similar
  $1/v^2$ dependence (e.g., the spin-dependent inelastic scattering
  considered in \cite{Kopp:2009qt}) but may feature a different
  dependence on $E_{nr}$. Furthermore, the shape test and the trivial
  bound can be applied for any particle physics model where the
  differential cross section factorizes as $X(v) Y(E_{nr})$, where $X$
  and $Y$ are arbitrary functions of $v$ and $E_{nr}$,
  respectively. An example where such a factorization is not possible
  in general are magnetic interactions, e.g.~\cite{Chang:2010en}. In
  such cases generalized methods as presented recently in
  \cite{DelNobile:2013cva} may be invoked.

\subsection*{Acknowledgements}

N.B., J.H.-G., and T.S.\ acknowledge support from the  European Union FP7  
ITN INVISIBLES (Marie Curie Actions, PITN-GA-2011-289442). J.H.-G.\ is  supported by the MICINN under the FPU program.
J.Z.\ was supported in part by the U.S. National Science Foundation under CAREER Grant PHY-1151392.

\bibliographystyle{my-h-physrev.bst}
\bibliography{./refs}

\begin{thebibliography}{10}

\bibitem{Goodman:1984dc}
M.~W. Goodman and E.~Witten,
\newblock {\em {Detectability of Certain Dark Matter Candidates}},
\newblock Phys.Rev. {\bf D31}, 3059 (1985).

\bibitem{Bernabei:2010mq}
DAMA/LIBRA Collaboration, R.~Bernabei {\em et~al.},
\newblock {\em {New results from DAMA/LIBRA}},
\newblock Eur.Phys.J. {\bf C67}, 39 (2010), 1002.1028.

\bibitem{Drukier:1986tm}
A.~K. Drukier, K.~Freese, and D.~N. Spergel,
\newblock {\em {Detecting Cold Dark Matter Candidates}},
\newblock Phys. Rev. {\bf D33}, 3495 (1986).

\bibitem{Freese:1987wu}
K.~Freese, J.~A. Frieman, and A.~Gould,
\newblock {\em {Signal Modulation in Cold Dark Matter Detection}},
\newblock Phys. Rev. {\bf D37}, 3388 (1988).

\bibitem{Ahmed:2009zw}
CDMS-II Collaboration, Z.~Ahmed {\em et~al.},
\newblock {\em {Dark Matter Search Results from the CDMS II Experiment}},
\newblock Science {\bf 327}, 1619 (2010), 0912.3592.

\bibitem{Ahmed:2010wy}
CDMS-II, Z.~Ahmed {\em et~al.},
\newblock {\em {Results from a Low-Energy Analysis of the CDMS II Germanium
  Data}},
\newblock Phys. Rev. Lett. {\bf 106}, 131302 (2011), 1011.2482.

\bibitem{Angle:2011th}
XENON10 Collaboration, J.~Angle {\em et~al.},
\newblock {\em {A search for light dark matter in XENON10 data}},
\newblock Phys.Rev.Lett. {\bf 107}, 051301 (2011), 1104.3088.

\bibitem{Aprile:2012nq}
XENON100 Collaboration, E.~Aprile {\em et~al.},
\newblock {\em {Dark Matter Results from 225 Live Days of XENON100 Data}},
\newblock Phys.Rev.Lett. {\bf 109}, 181301 (2012), 1207.5988.

\bibitem{TuckerSmith:2001hy}
D.~Tucker-Smith and N.~Weiner,
\newblock {\em {Inelastic dark matter}},
\newblock Phys. Rev. {\bf D64}, 043502 (2001), hep-ph/0101138.

\bibitem{Angloher:2011uu}
G.~Angloher {\em et~al.},
\newblock {\em {Results from 730 kg days of the CRESST-II Dark Matter Search}},
\newblock Eur.Phys.J. {\bf C72}, 1971 (2012), 1109.0702.

\bibitem{Kopp:2011yr}
J.~Kopp, T.~Schwetz, and J.~Zupan,
\newblock {\em {Light Dark Matter in the light of CRESST-II}},
\newblock JCAP {\bf 1203}, 001 (2012), 1110.2721.

\bibitem{Aprile:2011ts}
XENON100 Collaboration, E.~Aprile {\em et~al.},
\newblock {\em {Implications on Inelastic Dark Matter from 100 Live Days of
  XENON100 Data}},
\newblock Phys.Rev. {\bf D84}, 061101 (2011), 1104.3121.

\bibitem{Chang:2008gd}
S.~Chang, G.~D. Kribs, D.~Tucker-Smith, and N.~Weiner,
\newblock {\em {Inelastic Dark Matter in Light of DAMA/LIBRA}},
\newblock Phys.Rev. {\bf D79}, 043513 (2009), 0807.2250.

\bibitem{SchmidtHoberg:2009gn}
K.~Schmidt-Hoberg and M.~W. Winkler,
\newblock {\em {Improved Constraints on Inelastic Dark Matter}},
\newblock JCAP {\bf 0909}, 010 (2009), 0907.3940.

\bibitem{Kopp:2009qt}
J.~Kopp, T.~Schwetz, and J.~Zupan,
\newblock {\em {Global interpretation of direct Dark Matter searches after
  CDMS-II results}},
\newblock JCAP {\bf 1002}, 014 (2010), 0912.4264.

\bibitem{Arina:2012dr}
C.~Arina,
\newblock {\em {Chasing a consistent picture for dark matter direct searches}},
\newblock Phys.Rev. {\bf D86}, 123527 (2012), 1210.4011.

\bibitem{Kim:2012rza}
S.~Kim {\em et~al.},
\newblock {\em {New Limits on Interactions between Weakly Interacting Massive
  Particles and Nucleons Obtained with CsI(Tl) Crystal Detectors}},
\newblock Phys.Rev.Lett. {\bf 108}, 181301 (2012), 1204.2646.

\bibitem{MarchRussell:2008dy}
J.~March-Russell, C.~McCabe, and M.~McCullough,
\newblock {\em {Inelastic Dark Matter, Non-Standard Halos and the DAMA/LIBRA
  Results}},
\newblock JHEP {\bf 0905}, 071 (2009), 0812.1931.

\bibitem{Lisanti:2010qx}
M.~Lisanti, L.~E. Strigari, J.~G. Wacker, and R.~H. Wechsler,
\newblock {\em {The Dark Matter at the End of the Galaxy}},
\newblock Phys.Rev. {\bf D83}, 023519 (2011), 1010.4300.

\bibitem{Fox:2010bz}
P.~J. Fox, J.~Liu, and N.~Weiner,
\newblock {\em {Integrating Out Astrophysical Uncertainties}},
\newblock Phys.Rev. {\bf D83}, 103514 (2011), 1011.1915.

\bibitem{Fox:2010bu}
P.~J. Fox, G.~D. Kribs, and T.~M. Tait,
\newblock {\em {Interpreting Dark Matter Direct Detection Independently of the
  Local Velocity and Density Distribution}},
\newblock Phys.Rev. {\bf D83}, 034007 (2011), 1011.1910.

\bibitem{McCabe:2011sr}
C.~McCabe,
\newblock {\em {DAMA and CoGeNT without astrophysical uncertainties}},
\newblock Phys.Rev. {\bf D84}, 043525 (2011), 1107.0741.

\bibitem{Frandsen:2011gi}
M.~T. Frandsen, F.~Kahlhoefer, C.~McCabe, S.~Sarkar, and K.~Schmidt-Hoberg,
\newblock {\em {Resolving astrophysical uncertainties in dark matter direct
  detection}},
\newblock JCAP {\bf 1201}, 024 (2012), 1111.0292.

\bibitem{Gondolo:2012rs}
P.~Gondolo and G.~B. Gelmini,
\newblock {\em {Halo independent comparison of direct dark matter detection
  data}},
\newblock JCAP {\bf 1212}, 015 (2012), 1202.6359.

\bibitem{HerreroGarcia:2012fu}
J.~Herrero-Garcia, T.~Schwetz, and J.~Zupan,
\newblock {\em {Astrophysics independent bounds on the annual modulation of
  dark matter signals}},
\newblock Phys.Rev.Lett. {\bf 109}, 141301 (2012), 1205.0134.

\bibitem{Frandsen:2013cna}
M.~T. Frandsen, F.~Kahlhoefer, C.~McCabe, S.~Sarkar, and K.~Schmidt-Hoberg,
\newblock {\em {The unbearable lightness of being: CDMS versus XENON}},
\newblock (2013), 1304.6066.

\bibitem{DelNobile:2013cta}
E.~Del~Nobile, G.~B. Gelmini, P.~Gondolo, and J.-H. Huh,
\newblock {\em {Halo-independent analysis of direct detection data for light
  WIMPs}},
\newblock (2013), 1304.6183.

\bibitem{HerreroGarcia:2011aa}
J.~Herrero-Garcia, T.~Schwetz, and J.~Zupan,
\newblock {\em {On the annual modulation signal in dark matter direct
  detection}},
\newblock JCAP {\bf 1203}, 005 (2012), 1112.1627.

\bibitem{Bringmann:2009vf}
T.~Bringmann,
\newblock {\em {Particle Models and the Small-Scale Structure of Dark Matter}},
\newblock New J.Phys. {\bf 11}, 105027 (2009), 0903.0189.

\bibitem{Diemand:2005vz}
J.~Diemand, B.~Moore, and J.~Stadel,
\newblock {\em {Earth-mass dark-matter haloes as the first structures in the
  early Universe}},
\newblock Nature {\bf 433}, 389 (2005), astro-ph/0501589.

\bibitem{Kuhlen:2012fz}
M.~Kuhlen, M.~Lisanti, and D.~N. Spergel,
\newblock {\em {Direct Detection of Dark Matter Debris Flows}},
\newblock Phys.Rev. {\bf D86}, 063505 (2012), 1202.0007.

\bibitem{Freese:2012xd}
K.~Freese, M.~Lisanti, and C.~Savage,
\newblock {\em {Annual Modulation of Dark Matter: A Review}},
\newblock (2012), 1209.3339.

\bibitem{Gelmini:2000dm}
G.~Gelmini and P.~Gondolo,
\newblock {\em {WIMP annual modulation with opposite phase in Late-Infall halo
  models}},
\newblock Phys.Rev. {\bf D64}, 023504 (2001), hep-ph/0012315.

\bibitem{Green:2003yh}
A.~M. Green,
\newblock {\em {Effect of realistic astrophysical inputs on the phase and shape
  of the WIMP annual modulation signal}},
\newblock Phys. Rev. {\bf D68}, 023004 (2003), astro-ph/0304446.

\bibitem{BHSZ}
N.~Bozorgnia, J.~Herrero-Garcia, T.~Schwetz, and J.~Zupan,
\newblock {\em {in preparation}},
\newblock (2013).

\bibitem{Bozorgnia:2010xy}
N.~Bozorgnia, G.~B. Gelmini, and P.~Gondolo,
\newblock {\em {Channeling in direct dark matter detection I: channeling
  fraction in NaI (Tl) crystals}},
\newblock JCAP {\bf 1011}, 019 (2010), 1006.3110.

\bibitem{Chang:2010en}
S.~Chang, N.~Weiner, and I.~Yavin,
\newblock {\em {Magnetic Inelastic Dark Matter}},
\newblock Phys. Rev. {\bf D82}, 125011 (2010), 1007.4200.

\bibitem{DelNobile:2013cva}
E.~Del~Nobile, G.~Gelmini, P.~Gondolo, and J.-H. Huh,
\newblock {\em {Generalized Halo Independent Comparison of Direct Dark Matter
  Detection Data}},
\newblock (2013), 1306.5273.

\end{thebibliography}

\end{document}